\documentclass[english,aps,prper,reprint,showpacs,titlepage,longbibliography,nofootinbib,superscriptaddress]{revtex4-2}

\usepackage[T1]{fontenc}	
\usepackage[latin9]{inputenc}	
\usepackage{geometry}		
\usepackage{graphicx}
\usepackage[above,below]{placeins}	
\usepackage{times}
\usepackage{multirow}
\usepackage[dvipsnames]{xcolor}
\usepackage{subcaption}
\usepackage{titlesec}
\usepackage{ifthen}
\usepackage{makecell}
\newenvironment{iquote}
    {\vspace{-.43\baselineskip}\itshape\list{}{\leftmargin=0.15in\rightmargin=0.15in}%
    \item\relax}
    {\endlist\vspace{-.43\baselineskip}}

\newcommand{\ket}[1]{|#1\rangle}

\newcommand{\fracroot}[2]{\ifthenelse{#1=1}{\frac{1}{\sqrt{#2}}}{\sqrt{\frac{#1}{#2}}}}

\usepackage{makecell}

\newcommand{\who}[1]{\textbf{#1\emph{:}}}

\newif\ifshowtimestamp
\showtimestampfalse

\usepackage{hyperref}  
\hypersetup{colorlinks=true,urlcolor=blue,citecolor=blue,linkcolor=blue}   
\usepackage{enumitem}        
\setlist{nosep}                 

\usepackage{mathtools}
\usepackage{amssymb}
\usepackage{amsmath}
\usepackage{amsthm}
\usepackage[font=small]{caption}
\usepackage{subcaption}
\usepackage{bbold}
\usepackage{comment}
\usepackage{tabularx}
\usepackage{array}
\usepackage{censor}

\usepackage[dvipsnames]{xcolor}
\usepackage{footmisc}

\usepackage[sort&compress]{natbib}

\titlespacing{\section}{0pt}{10pt plus 2pt minus 1pt}{10pt plus 2pt minus 1pt}

\titlespacing{\subsection}{0pt}{10pt plus 2pt minus 1pt}{10pt plus 2pt minus 1pt}

\begin{document}
\setlist[itemize]{leftmargin=20pt}
\begin{titlepage}

\title{Computer-generated QIS tutorial feedback is valued by students, but does not replicate in-class collaboration}
\author{Josephine C.\ Meyer}
\affiliation{Department of Physics, California State University - San Marcos, San Marcos, CA 92096, US}
\affiliation{Department of Physics \& Astronomy, George Mason University, Fairfax, VA 22030, US}
\author{Steven J.\ Pollock}
\author{Bethany R.\ Wilcox}
\affiliation{Department of Physics, University of Colorado - Boulder, Boulder, CO 80309, US}
\author{Gina Passante}
\affiliation{Department of Physics, California State University - Fullerton, Fullerton, CA 92831, US}

\begin{abstract}
PER has consistently demonstrated the effectiveness of small-group tutorials in helping students develop conceptual understanding and fluency, but instructor uptake is limited by resource constraints.
To test the effectiveness of out-of-class tutorials using computer-generated feedback as an instructor-friendly alternative, we conducted think-aloud interviews with students in a quantum computing course who were randomly assigned to either a traditional validated small-group, pencil-and-paper tutorial on tensor products, or a solo computerized adaptation thereof.
We found that while the computer-generated feedback was broadly considered useful by students, 
student engagement patterns were markedly different in the solo setting, with students demonstrating reluctance to use the interface's built-in help features and tending to internalize failure in unproductive ways counter to our intention of a formative learning environment. We discuss implications for curriculum design and directions for future research that may help to answer the longstanding question in PER of \textit{why} tutorials work so well.

\clearpage
\end{abstract}

\maketitle
\end{titlepage}

\section{Introduction and background}

Small-group tutorials have been consistently shown to be an effective pedagogical strategy at all levels of physics education \cite{McDermott:1998,DeVore:2014thesis,Passante:2015b,Xue:2016,Porter:2020,Pollock:2023b}, with recent work pointing to tutorials and similar long-form small group activities as a common feature among the active-learning physics courses that exhibit the highest learning gains \cite{Sundstrom:2025c}. 
Development of tutorials remains an active area for PER, including for emerging domains such as quantum information science (QIS) \cite{AcePhysics,Hu:2024b,Thacker:2025b}.

Less clear to the community is the specific mechanisms by which tutorials produce strong and consistent learning gains. Some studies have pointed to factors in the design of tutorials themselves, such as the grounding in theories of learning, iterative design process, and ``revisiting'' or contrasting case strategies \cite{VonKorff:2016,Kuo:2016}. Other studies have highlighted the central role of Socratic instructor-student dialogue \cite{Koenig:2004,Slezak:2011} and peer interactions \cite{Conlin:2007,Leinonen:2017,Cervantes:2022} within tutorial environments. 
Replication studies further complicate the picture, with several studies showing promising results across contexts (e.g.~\cite{Finkelstein:2005a,Mauk:2005,Keller:2006,Koenig:2007,Benegas:2014a,Kryjevskaia:2014a}) while others showed heavy variation across instructors \cite{Pollock:2008} or failed to meaningfully improve on traditional lecture \cite{Sabella:2002,Riegler:2016}.

Yet despite growing awareness of the effectiveness of tutorials and other research-based instructional strategies (RBIs), uptake among physics educators remains stubbornly low \cite{Dancy:2024}. Several studies have investigated causes for faculty resistance to RBIs and barriers to implementation in physics, with classroom layout and time constraints (both class time -- a phenomenon dubbed ``tyranny of content'' in biology education \cite{Petersen:2020} -- and preparation time) as underlying themes \cite{Henderson:2007a,Michael:2007,Dancy:2010,Turpen:2016,Shadle:2017,Apkarian:2021,Carroll:2023}.%

\begin{figure}
    \centering
    \includegraphics[width=\linewidth, trim = 40pt 40pt 40pt 100pt, clip=true]{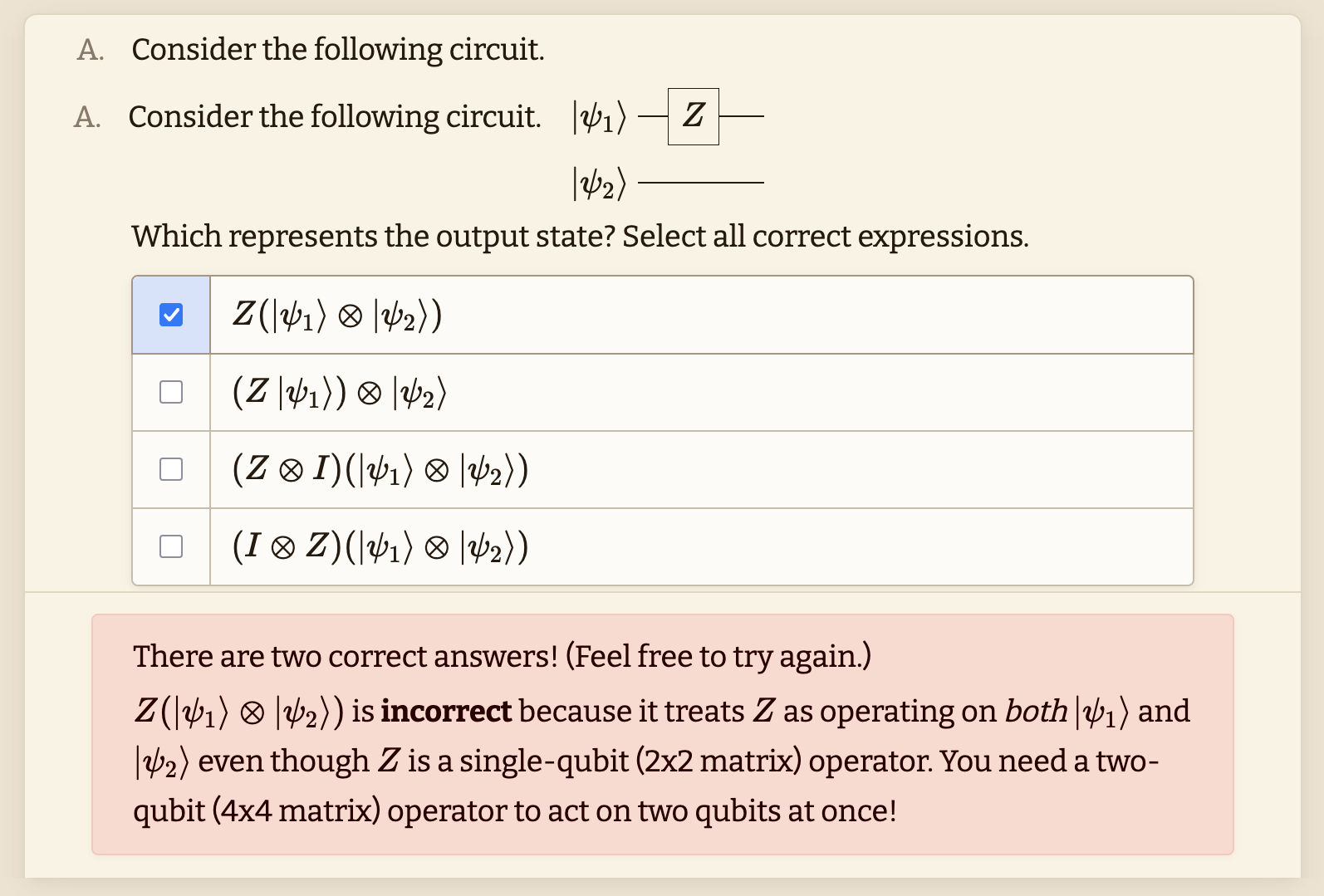}
    \caption{Screenshot of the AceQIS interface for the tutorial used in this study (tightened for space). The dynamic feedback box provides detailed feedback specific to the student's response.}
    \label{fig:screenshot}
\end{figure}

With the belief that an imperfect instantiation of tutorials is almost certainly better than the \textit{de facto} baseline of no tutorials at all, our team developed AcePhysics \cite{AcePhysics}, an online platform using computerized feedback to mimic the Socratic style of questioning and feedback students may receive from instructors or peers in the classroom,\footnote{Students receive real-time feedback in response to closed-form answers, mimicking the feedback a student may receive from a TA. Some problems offer optional hints. While recent studies have explored AI-enabled feedback (e.g.~\cite{El-Adawy:2024,Krupp:2025,Dange:2026}), we intentionally designed AcePhysics \textit{not} to rely on AI, in part because at the time of development, available AI models proved unable to replicate Socratic dialogue central to tutorial implementation.} while allowing instructors to assign the tutorial as a ``plug-and-play'' out-of-class task \cite{Corsiglia:2023thesis} that would not consume class time or require a particular classroom layout. 
Each AcePhysics tutorial is based on a pencil-and-paper tutorial developed by our group for use in quantum mechanics and/or QIS courses \cite{Pollock:2023b}. 
The structure of each AcePhysics tutorial adheres closely to the structure of the pen-and-paper tutorial on which it is based, with adaptations as necessary to fit the virtual medium (see Fig.~\ref{fig:screenshot}).

Prior work provided preliminary evidence of the effectiveness of AcePhysics tutorials for remote group collaboration during emergency remote teaching \cite{Corsiglia:2023thesis} while noting several limitations \cite{Corsiglia:2022b,Corsiglia:2023thesis} and exploring group dynamics \cite{Cervantes:2022}. However, until now, AcePhysics has seen only limited testing under the format's original intent -- faculty assigning tutorials as individual homework problems outside of class. 
Though two prior studies with computer-based tutorials in intro physics showed underwhelming results \cite{Slezak:2011,DeVore:2017}, it was unclear how these findings generalized to the upper-division.

\section{Methodology}

We conducted think-aloud interviews with seven students in an interdisciplinary upper-division quantum computing course at a large R1 university in spring 2026. Students were quasi-randomly assigned\footnote{Initial assignments utilized stratified random sampling by major. $N=9$ students signed up for interviews and were assigned 6 to the traditional group tutorial (2 groups of 3) and 3 to the individual AcePhysics tutorial. When 2 students in the same 3-person tutorial group no-showed, the remaining student was reassigned to solo AcePhysics.} to complete a tutorial on basics of tensor products (a key QIS skill) under one of two conditions to compare student engagement with each:

\begin{itemize}
    \item \textbf{Traditional tutorial:} Students worked in groups of 3 to complete the tutorial on paper in-person. Interviewer JCM, 
    who had previously served as a TA for the course, simulated the interaction of a course TA.
    \item \textbf{AcePhysics tutorial:} Individuals completed the online tutorial on their personal device, similar to how it might be administered as homework. JCM 
    was present to answer logistical/UI questions and guide the think-aloud process, but refrained from giving direct feedback or content-related hints unless absolutely necessary to allow the student to continue with the tutorial.
\end{itemize}

Interviews were conducted near the end of the semester, after all relevant material had been introduced in class. Students were incentivized with a \$30 gift card for participation. Demographics of the participants are given in Table~\ref{tab:demographics}. Though sensitive demographics (e.g.\ race, gender) were not collected due to small $N$, the sample appeared to reflect the primarily white and male course population.

\begin{table}[tb]
    \centering
    \begin{tabular}{c c c c c}
        \hline\hline
         \thead{Cond.} & \thead{Student}& \thead{Major(s)} & \thead{Minor(s)} & 
        \thead{Pretest}\vspace{-3pt}\\ 
         \hline \vspace{-9pt}\\

          \multirow{4}{*}{\rotatebox[origin=c]{90}{\makecell{Individual \\ AcePhysics}}} & Alan & Applied math & CS, EE & 3.5 / 4\\

         & Bryce & ECE & CS, QE & 2.5 / 4 \\
         & Carlos & CS & -- & 1 / 4 \\

         & Diana & Astrophysics & QE & 1.5 / 4 \\

         \hline \vspace{-9pt}\\ 

         \multirow{3}{*}{\rotatebox{90}{\makecell{Group \\ (Paper)}}} & Xavier  & ECE & CS, QE & 3.5 / 4\\

         & Yuri  & Physics, math & CS, QE & 2 / 4 \\

         & Zach  & Eng.\ physics & QE & 4 / 4
         \\ 

        \hline\hline
    \end{tabular}
    \caption{Academic major(s)/minor(s), pretest score, and tested tutorial condition for the 7 student interviewees in this study. CS = computer science, ECE = electrical and computer engineering, EE = electrical engineering, QE = quantum engineering}
    \label{tab:demographics}
\end{table}

\begin{figure}
    \centering
    \includegraphics[width=0.6\linewidth, trim = 20pt 10pt 10pt 0pt]{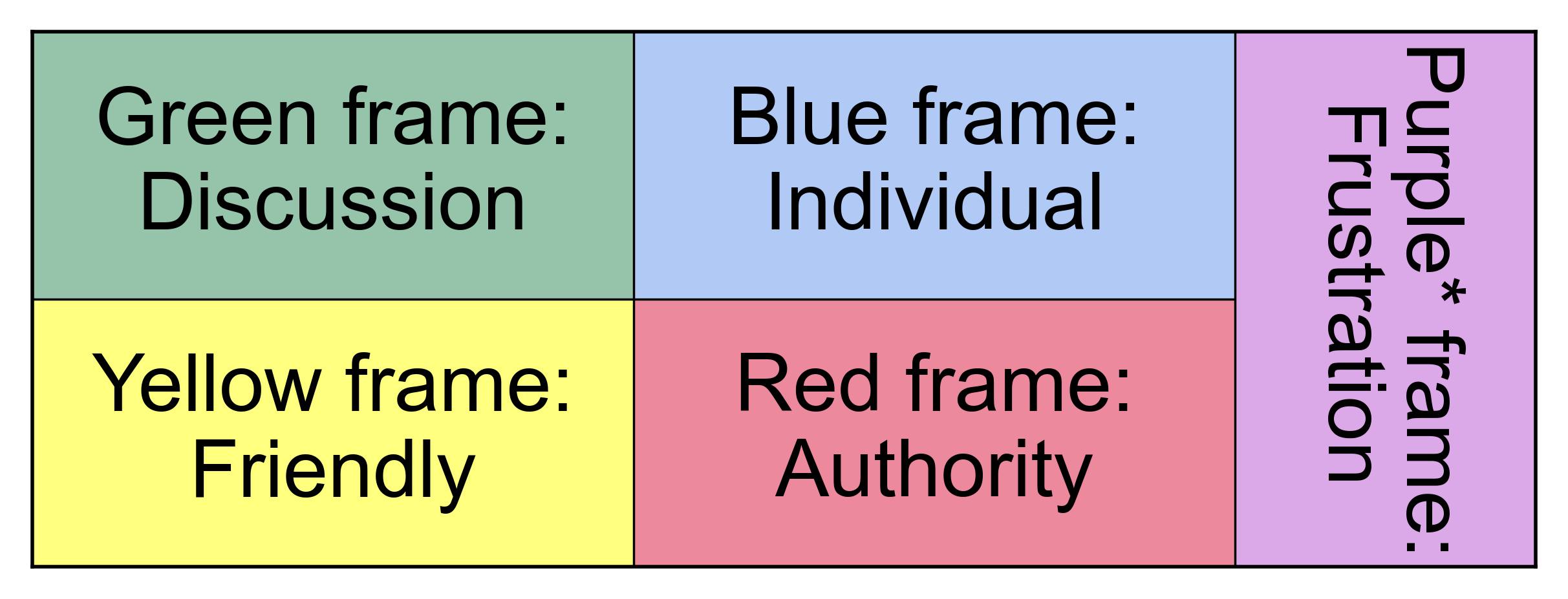}
    \caption{Color frames for tutorial work. See Refs.~\cite{Scherr:2009,Cervantes:2022} for definitions and coding schema. *Purple frame newly proposed.}
    \label{fig:colorframes}
\end{figure}

Interviews were scheduled for 1 hour (reflecting the 50 min design length of the original pencil-and-paper tutorial), and students were prompted to think aloud throughout the activity. 
Interviews were video-recorded, and audio was transcribed using Otter.ai with human verification. Paper tutorial worksheets as well as scratch work generated during the AcePhysics tutorials were collected as artifacts. Transcripts and artifacts were open-coded for themes using Dedoose, with the dual intent of probing student engagement with the AcePhysics vs.\ paper tutorials (this article) and student understanding of tensor products (to be published separately). This article focuses on the following research question: \textit{How does student interaction with a solo AcePhysics tutorial (i.e.\ under conditions similar to a homework problem) compare to the same tutorial under classroom-like group conditions?}

For our analysis, we focused on those activity segments where the features of the AcePhysics interface were assumed most salient: specifically, those times when students sought, received, and/or were voluntarily offered guidance or feedback beyond the text of the tutorial. Likewise, we focused on individual or group behavior following an acknowledged error or mistake. 
We also analyzed the videotaped tutorials using Scherr's value-neutral color frames \cite{Scherr:2009} as adapted by Cervantes \textit{et al.}~\cite{Cervantes:2022} (Fig.~\ref{fig:colorframes}) to examine engagement patterns.

\section{Results and discussion}

In this section, we report and comment on key outcomes for each tested tutorial modality (pencil-and-paper group work vs.\ solo AcePhysics). 
In Sec.~\ref{sec:selfreport} we discuss students' self-reported feedback about the tutorial, which was highly positive for both modalities. However, students report not wanting to engage with the hints and similar student-centered features of the AcePhysics UI, potentially limiting the formative pedagogical value of the interface and signaling a possible disconnect between the developers' and students' values and expectations for AcePhysics. Then, in Sec.~\ref{sec:results-colorframes}, we analyze student engagement patterns using the color frames of Scherr \cite{Scherr:2009} and Cervantes \textit{et al.}~\cite{Cervantes:2022}, focusing on students' behavior following identified mistakes. We contrast the group tutorial, in which the Yellow frame appeared to defuse psychological tension and reestablish an atmosphere of collegial equality following a mistake, with the individual tutorial, in which students tended to get stuck in a ``Purple frame'' (not observed in the group tutorial or in prior work) of self-criticism counter to the formative intent.

\subsection{Student self-report feedback and UI interaction}
\label{sec:selfreport}

While students had suggestions to improve the UI (e.g.\ adding an equation editor), overall impressions of the tutorial experience were overwhelmingly positive for both modalities. Interviewees expressed a desire to see activities similar to the AceQIS tutorial on their homework, and those with prior experience with tutorials typically felt AcePhysics was a satisfactory emulation. Diana desired more activities like this in her course, even if added to existing homework loads:

\begin{iquote}
    \who{Diana} I don't see it necessarily being better working in a group or alone ... 
    I would do this even ... on top of everything else that I have to do. Any help would~be~great.
\end{iquote}

\noindent While Carlos felt solo AcePhysics a poor substitute for tutorial, he and Diana suggested using it to supplement or replace the prelecture ``preflight'' quizzes in the course (testable in future work).
Alan preferred human-human TA interaction for the ability to rephrase if he did not understand the first time, but added that AcePhysics was much preferable to ``when you're at home and struggling'' on homeworks alone.

However, when students were asked to reflect on their engagement with the growth-focused elements of the AcePhysics interface, responses suggested students still interacted with the AcePhysics tutorials from an answer-correctness mindset (possibly transferred from graded homeworks) rather than as a formative activity for growth as intended by the developers. For instance, Diana reflected that she did not click the hints even when she likely would have benefited from them, because she felt the material was ``pretty basic'' and therefore she shouldn't need them:

\begin{iquote}
    \who{Diana} I just really didn't want to use the hints because ... this is pretty basic starting stuff ... so I didn't want to click the hints unless I was really struggling with them.
\end{iquote}
After the activity, Carlos also reflected on his nonuse of hints:

\begin{iquote}
    \who{Carlos} Whenever I do other assignments where hints are also offered, I feel like I never use them ... I just didn't think to look at the hint.
\end{iquote}
Carlos elaborated that he tended to work on the problem by himself and then type in the answer, looking only at the prompt and answer box, without even considering whether other support might be available on the screen. His response suggests that the mere availability of friendly hints and feedback on the online interface is insufficient to use them, if students are used to a ``read, calculate, input, submit'' workflow for online homework problems. Regardless, Carlos and Diana's nonuse of the hints ran counter to both (a) the AcePhysics developers' intention of a friendly, formative atmosphere, and (b) the dynamics observed in the actual in-person tutorial, where help was readily both offered and solicited until all students agreed upon an approach and answer.

Overall, features like hints seemed to be only as useful as students allowed them to be. Bryce, who \textit{did} make ample use of the hints without prompting, commented that they may be  useful for students truly seeking to learn the material but not for students aiming to speed through the assignment:
\begin{iquote}
    \who{Bryce} The hints are a good substitute for a TA. I think if someone were coming at it with the wrong intentions ... they could get nothing out of this ... The hints only helped if you were trying to do it.
\end{iquote}
Bryce demonstrated keen metacognition skills throughout the interview. His comment insinuates that the AcePhysics platform may be useful for students with high levels of self-awareness and orientation toward learning, but may not help students with weaker metacognitive skills learn to build them (which explicit interaction with TAs and/or peers may scaffold). This result may also help explain DeVore \textit{et al.}'s findings that lower-achieving students (whom we as instructors might wish to prioritize reaching) benefit the least from self-paced formative activities like AcePhysics \cite{DeVore:2017}.

One such metacognitive skill is discerning between productive and unproductive struggle. In in-person tutorials, unproductive struggle can be interrupted by a TA or other students. AcePhysics attempts to provide some of these cues via embedded timing and reminders to move on if stuck, but they will only be useful if students have sufficient awareness to recognize they need them. Tellingly, both Carlos and Diana closed a system dialogue box advising them not to spend more than 5 minutes on the pretest warm-up and continued to persevere fruitlessly; Carlos remarked he hadn't even read it.

\subsection{Color frames and student reactions to errors}
\label{sec:results-colorframes}

The group pencil-and-paper tutorial showed a lively mixture of the Green, Blue, and Yellow frames. The Yellow frame largely coexisted with the Green and Blue frames, similar to that reported for one group in Ref.~\cite{Cervantes:2022}, with conviviality appearing not to distract from completion of the tutorial, and possibly even to increase motivation. The Red frame was seldom observed, potentially because material was review.

As expected, the Green and Yellow frames were essentially absent in the solo AcePhysics tutorials. Instead, students tended to switch between the Blue frame (individual problem-solving) and Red frame (checking answers, or occasionally asking the facilitator for help/clarification where the computer's feedback was ambiguous or insufficient).

We also observed evidence of a new ``Purple'' frame within the solo tutorials, which we tentatively describe as the ``frustration'' frame. The Purple frame was characterized by negative self-talk. It often appeared to be activated by frustration (e.g.\ receiving feedback that an answer was incorrect, particularly when the mistake was minor) and/or boredom (e.g.\ tedious calculations). In contrast, the Purple frame was not seen to a meaningful extent in the group discussion.

The most notable difference came in how students responded when they realized (either from AcePhysics feedback or a peer) that they had made a mistake -- particularly a minor mechanical mistake such as an arithmetic or sign error. In the pencil-and-paper group tutorial, students tended to shift into the Green or Yellow frame following such mistakes and move on quickly. Sometimes mistakes even became humorous opportunities for bonding, as in the following snippet:
\begin{iquote}
    \who{Zach} Wait, no ... $\ket{01}$ is the second state in the vector.\vspace{2pt}\\
    \who{Yuri} Oh my g-d I thought it was $\ket{10}$!\vspace{2pt}\\
    \who{Xavier} I did that too... [chuckles sheepishly]\vspace{2pt}\\
    \who{Zach} I'm glad that they didn't mess with binary [ordering] for the vector notation. \hspace{3pt} \who{Yuri} Yeah.
\end{iquote}
Friendly discourse about notation conventions (Yellow frame) was a common thread throughout the entire interview, particularly while students were working on individual calculations (Blue frame). This snippet of dialogue, particularly Zach's follow-up comment, functioned to firmly place all 3 students back on an equal footing after Yuri and Xavier's potentially embarrassing mistake. Yuri and Xavier were able to bond over having made the same mistake, and Zach attributed his own correct recollection to the merits of the notation (a reference back to the ongoing Yellow frame banter).

However, students completing the AcePhysics solo tutorial tended to double down on negative self-talk to the exclusion of productive reflection. For instance, Diana made a decimal conversion error ($2/5\rightarrow0.2$) and was surprised to see AcePhysics mark her answer as wrong. Interviewer JCM pointed out it was just a minor arithmetic error and laughed it off. When she got to the bottom of the answers page and was asked to reflect on her learnings, however, Diana moved to the Purple frame, still focused on her arithmetic error:
\begin{iquote}
    \who{Diana} Ugh, okay. Well, I'm just going to write ``2/5 is actually .4'' [in the box].
\end{iquote}
Diana quickly moved on to the next page, still appearing embarrassed, with no reflection on the actual substantive content. 

Earlier, Diana 
had correctly typed the output state of a circuit as ``{\small$\ket-\ket1$},'' where {\small$\ket-$=$\fracroot{1}{2}{\ket0-\ket1}$} is common shorthand. Since free-form text is challenging to autograde, AcePhysics asked her to manually compare her answer with a reference, open-endedly given as ``{\small{$\fracroot{1}{2}\ket{01} - \fracroot{1}{2}\ket{11}$}}, {\small{$\fracroot{1}{2}(\ket0-\ket1)\otimes\ket1$}} or, perhaps even: {\small{$-\fracroot{1}{2}(\ket1-\ket0)\otimes\ket1$}}.''
In presenting 3 valid forms of the same answer, we aimed to focus students' attention on structural rather than superficial features. Yet when prompted by AcePhysics to reflect on her learning, despite acknowledging her answer's equivalency, Diana immediately critiqued it for being (in her opinion) aesthetically inferior:
\begin{iquote}
    \who{Diana} I learned that ... the actual things [standard basis vectors shown in model solution] look a little better than this [her answer].
\end{iquote}

\noindent Taken together, we view these two cases as evidence of an answer-matching mindset (faithfully seeking to reproduce the computer's ``right'' answer at all costs) rather than the intended formative learning mindset emphasizing the conceptual reasoning pathways necessary to arrive at that answer. Meanwhile, Carlos misremembered the matrix for the $Z$ operator. When 
alerted to this error, he turned the critique inward, remarking, ``My brain is not doing well today.''

Similar to the Yellow frame, the Purple frame frequently involved joking. However, humor in the Purple frame tended to be expressly self-deprecating, while humor in the Yellow frame in group work was built around a shared sense of camaraderie. It is therefore unclear whether the Purple frame is a fundamentally different modality of engagement from the Yellow frame or simply represents the same underlying impulses as the Yellow frame without the convivial outlet of a group. It is also possible the Purple frame is an artifact of entrenched associations with penalties in graded homework, or that the think-aloud format is simply surfacing negative self-talk that exists but is not otherwise voiced during groupwork.

Taken together, our observations indicate it may be the \textit{sociocultural context of collaborative group work itself} -- not the physics content, structure, or feedback tone (all essentially identical between the two conditions) -- may play an important role in the high learning gains of in-person tutorials. The Purple frame suggests a possible explanation for the previously-documented lukewarm success of solo tutorials with computer-generated feedback: perhaps the sociocultural context of group tutorials mitigates the tendency to turn frustration inward (particularly for lower-performing students). The Yellow frame in groupwork settings might serve as an important outlet to sublimate feelings of frustration which could otherwise block learning. For instance, for Xavier, Yuri, and Zach, the ongoing Yellow frame banter about notation conventions appeared to serve as a level field they could all return to following any student's mistake, thus ensuring no one internalized a feeling of being less than (despite Yuri's markedly lower pretest score). The non-observation of the Purple frame in group AcePhysics tutorials under emergency remote teaching (Ref.~\cite{Cervantes:2022}) lends credence to this hypothesis.

\section{Conclusions}
\label{sec:conclusion}

Our observations of the AcePhysics tutorial were encouraging in some respects. We found that students generally viewed the AcePhysics tutorial positively as a learning tool and showed evidence of learning during the activity. For some students like Bryce with strong self-motivation and metacognitive skills, AcePhysics indeed appeared to function comparably effectively to the paper tutorial. For most students, the AcePhysics tutorial presented useful practice when used alongside lecture, which interviewees consistently stated left them without a solid understanding of tensor products.

At the same time, despite the developers' attempts to expressly frame AcePhysics tutorials as a learning opportunity, we observed that some students (particularly Carlos and Diana -- the students in the AcePhysics condition with the lowest pretest scores, who would ideally stand to gain the most) appeared to interact with the tutorial primarily from an answer-matching mentality to the exclusion of genuine formative reflection, echoing Refs.~\cite{Slezak:2011,DeVore:2017}. Student engagement patterns were markedly different from developers' intentions \cite{Corsiglia:2023thesis}, themselves rooted in the learning theory view of tutorials \cite{McDermott:2021}, leave us skeptical that the robust learning gains consistently observed for group tutorials would transfer to solo AcePhysics. (Large-scale quantitative data would be needed to definitively affirm or reject this hypothesis.) 

While our study has too small an $N$ to be conclusive (perhaps other students feel less comfortable making mistakes in front of peers than a computer), our observations do suggest potential explanations for \textit{why} peer interaction is so essential for tutorials. In particular, we see evidence that students responded differently to failure and frustration in the individual and group settings. Despite the friendly tone of the computer's responses, students in the solo case tended to turn toward negativity (Purple frame) following mistakes rather than using them as an opportunity to learn. 
Whatever the mechanism for the differential performance of the solo AcePhysics vs.\ group paper tutorials, our observations suggest an important limitation of AcePhysics as a solo tutorial environment (in line with Ref.~\cite{Corsiglia:2023thesis}). \textit{AcePhysics is not a substitute for in-person tutorials, even though it duplicates several aspects of them.} The sociocultural experience of the tutorial setting likely cannot be replicated solo. Still, from self-report feedback, AcePhysics appears clearly better than no tutorial at all.

\subsection{Limitations and alternate interpretations}

A key limitation of this study is that 1:1 interviews are an imperfect reflection of the solo homework environment. The interviewer (even if largely silent) and distraction-reduced clinical interview environment will not be present during a typical homework session. In Ref.~\cite{DeVore:2017}, students saw much greater learning gains with solo tutorial under clinical interview conditions than they did when assigned as homework problem, attributed in part to distractions and multitasking. Therefore, we might anticipate actual differences in student engagement between group pencil-and-paper vs.\ solo AcePhysics tutorials to be \textit{greater} in the field than in our study. 

On the other hand, UI-specific issues (such as student underuse of hints and discomfort with ASCII input of equations) might be expected to diminish with repeated exposure to the interface throughout the class. Notably, groupwork activities had previously been implemented in the quantum computing course from which students were drawn for this study, so students may have already established the epistemic and sociocultural norms to function productively in small-group environments. Perhaps repeated use of the AcePhysics platform throughout the course is necessary to realize the full benefits of the platform, similar to the threshold effect for traditional tutorials observed in Ref.~\cite{Slezak:2011}. Likewise, the researcher's presence and/or the act of thinking aloud may have contributed to a self-critical attitude; regardless, the observed frustration was stark enough to likely represent a real effect.

\section{Acknowledgments}

We thank Giaco Corsiglia and Jonan-Rohi Plueger for their work implementing the AcePhysics version of this tutorial. This work is funded by NSF grants No.\ PHY-2011958 and PHY-2012147.

\clearpage
\bibliography{PERC2026}

\end{document}